\RequirePackage{fix-cm}
\documentclass[smallextended]{svjour3}       % onecolumn (second format)
\smartqed  % flush right qed marks, e.g. at end of proof
\usepackage{graphicx}
\begin{document}

\title{Measurement of Permanent Electric Dipole
Moments of Charged Hadrons in Storage Rings%\thanks{Grants or other notes
}


\author{J\"org Pretz \\
on behalf of the {\sc JEDI} collaboration %       \and
      }


\institute{J\"org Pretz \at
              III. Physikalisches Institut \\
Physikzentrum 26C 212 \\
RWTH Aachen\\
52056 Aachen\\
              Tel.: +49 241 80-27306\\
              \email{pretz@physik.rwth-aachen.de}           %  \\
}

\date{Received: date / Accepted: date}

\maketitle

\begin{abstract}
Permanent Electric Dipole Moments (EDMs) of elementary particles violate
two fundamental symmetries: time reversal invariance ($\mathcal{T}$) and parity ($\mathcal{P}$). Assuming the $\mathcal{CPT}$ theorem this implies
$\mathcal{CP}$ violation. The $\mathcal{CP}$ violation of the Standard Model is 
orders of magnitude too small to be observed experimentally
in EDMs in the foreseeable future. It is also way too small 
to explain the asymmetry in abundance of matter and anti-matter
in our universe.
Hence, other mechanisms of $\mathcal{CP}$ violation outside the realm of the Standard Model 
are searched for and could result in measurable EDMs. 

Up to now most of the EDM measurements were done with neutral particles.
With new techniques it is now possible to perform
dedicated EDM experiments with charged hadrons at storage rings
where polarized particles are exposed to an electric field.
If an EDM exists the spin vector will experience a torque 
resulting in change of the origi\-nal spin direction which can be
determined with the help of a polarimeter.
Although the principle of the measurement is simple,
the smallness of the  expected effect makes this a challenging experiment
requiring new developments in various experimental areas.

Complementary efforts to measure EDMs of proton, deuteron and light nuc\-lei are
pursued at Brookhaven National Laboratory and at
Forschungszentrum J\"ulich with an ultimate
goal to reach a sensitivity of $10^{-29}e\cdot$cm. 
\end{abstract}

\section{Introduction \& Motivation}\label{intro}
One of the great mysteries in particle physics is the dominance
of matter over anti-matter in our Universe.
The net baryon number is~\cite{cmb}%wmap collaboration, \Omega_B = 0.0456\pm0.0016
\[
\frac{n_B-n_{\bar B}}{n_\gamma} = \left(6.1^{+0.3}_{-0.2} \cdot 10^{-10}  \right) \,.
\]
In the Standard Model (SM) this ratio
is expected to be on the order of $10^{-18}$.
In 1967 Sakharov~\cite{Sak67} formulated three prerequisites for baryogenesis.
One of these is the combined violation of the charge and parity, $\mathcal{CP}$, symmetry.
New $\mathcal{C}\mathcal{P}$ violating sources outside the realm of the SM
are clearly needed to explain this discrepancy of eight orders of magnitude.

For non-degenerate systems like for example elementary particles, including hadrons, an electric dipole moment is only possible if 
the two fundamental symmetries, parity $\mathcal{P}$ and time reversal invariance $\mathcal{T}$
are violated. Using the $\mathcal{CPT}$ theorem a violation of $\mathcal{T}$
is equivalent to  $\mathcal{C}\mathcal{P}$ violation.

$\mathcal{CP}$ violation is present in the Standard Model in two places.
In the so called $\theta$ term of QCD which could in principle range from 0 to 2$\pi$.
The fact that no hadronic EDM has been found up to now limits $\theta$ to 
an unnaturally small number of $10^{-10}$. This is called the strong QCD problem.
The second source is the 
complex phase parameter of the Cabibbo-Kobayashi-Maskawa matrix
which would result in EDMs of the order of $10^{-31}$ to $10^{-32} e\cdot $cm for hadrons, much below 
experimental sensitivity reachable in the near future, whereas most of the extension
of the Standard Model predict EDMs which are the range of future experiments~\cite{Pendlebury:2000an,sm_edm}.

%Thus the discovery of an EDM would be a clear signal for physics beyond
%the SM. 
Tab.~\ref{tab:had_edms} shows current limits on hadron EDMs.
There is no direct measurement of charged hadron EDMs reaching the sensitivity of the neutron measurement.
The measurement of a single hadron EDM cannot decide on the source of $\mathcal{C}\mathcal{P}$ violation
(e.g. $\theta$-term, beyond SM).
It is thus mandatory to measure EDMs of various species of particles\cite{deVries:2011an,Bsaisou:2012rg,vries,ritz,Guo:2012vf}.
Methods to determine charged hadron EDMs with a sensitivity of $10^{-29}e
\cdot$cm will be presented in the following. This subject is also addressed in
\cite{Onderwater:2012me}.

\begin{center}
\begin{table}[b!]
 \begin{tabular}{|l|c|c|c|}
\hline
Particle/Atom  &  Current Limit/$e\cdot$cm   & Ref. \\
\hline
 Neutron  &  $<3 \cdot 10^{-26} \quad (90\% CL)$   &  \cite{n_edm}  \\
$^{199}$Hg  &  $<3.1 \cdot 10^{-29} \quad (95\% CL)$  &     \\
%%% \rowcolor{blue!20} $^{129}$Xe & $< 6 \cdot 10^{27}$         \\
\hspace{3mm}$\rightarrow$ Proton &  $<7.9 \cdot 10^{-25}$ & \cite{hg_edm}    \\
 Deuteron &   -       &                \\
 $^3$He   &   -       &     \\
$\Lambda$ &  $(-3.0 \pm 7.4) \times 10^{-17}$  & ~\cite{Pondrom:1981gu} \\
\hline
\end{tabular}
\caption{Current limits of hadron EDMs.\label{tab:had_edms}}
\end{table}
\end{center}

\section{Principle of the Measurement}\label{princ}
The principle of every EDM measurement (be it atom, molecule, charged particle, \dots)
is the interaction of an electric field $\vec E$ with the dipole moment $\vec d$
of the particle.
Since the spin is the only vector of an elementary particle defining a direction,
the EDM has to be (anti-)parallel to the spin vector.
Thus under the influence of an electric field the spin vector $\vec S$ gets 
tilted (with respect to the momentum vector) according to
\begin{equation}
\frac{{\rm d}\vec S }{{\rm d} t} = \vec d  \times \vec E^* \, .
\end{equation}
Here, $\vec E^*$ denotes the electric field in the particle rest frame.

\begin{figure}[hp]
  \includegraphics[width=0.9\textwidth]{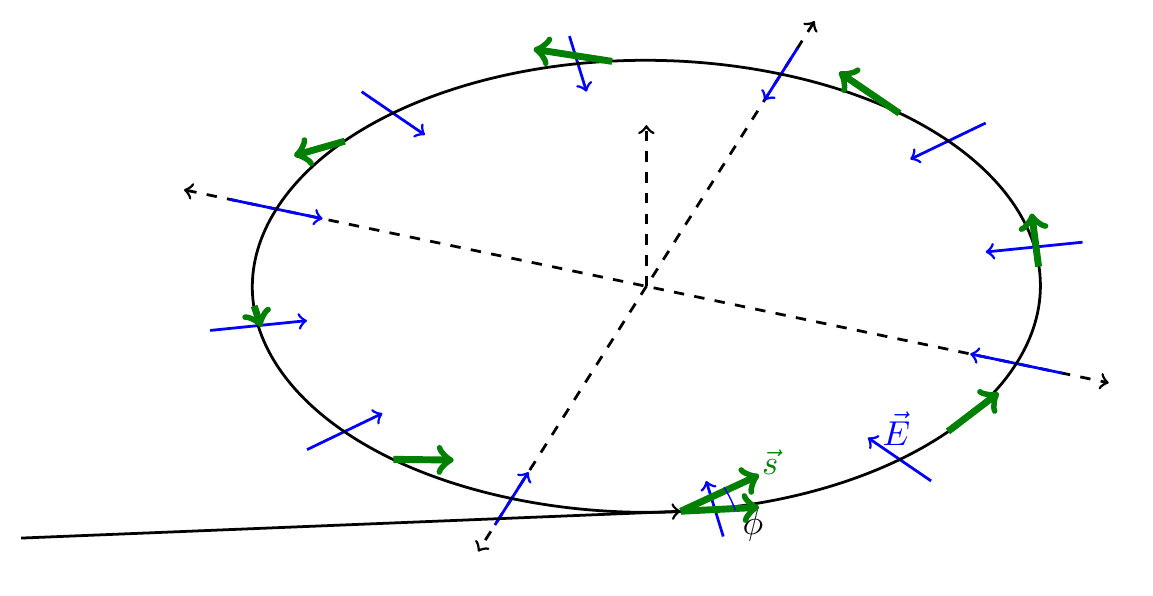}
\caption{Principle of an EDM measurement in a storage ring.
Longitudinally polarized particles enter a storage ring.
A radial electric field serves as a guiding field. An EDM will tilt the spin 
in the vertical direction. This vertical polarization can be measured with the
help of a polarimeter (not shown in the Figure).}
\label{fig:princ}       % Give a unique label
\end{figure}

A generic EDM measurement for a charged particle could thus 
look as indicated in Fig.~\ref{fig:princ},~\cite{Farley:2003wt,yannis1}
\footnote{There is one (parasitic) measurement for muons~\cite{Bennett:2008dy}
using this technique.}.
Longitudinally polarized particles enter a storage ring.
A radial electric field serves as a guiding field. An EDM will tilt the spin 
in the vertical direction. This vertical polarization component can be measured with the
help of a polarimeter.

More generally in presence of electric and magnetic fields, and
considering that particles also posses a magnetic moment 
$\vec \mu= 2(G+1) \frac{e \hbar}{2m} \vec S$ ($G=(g-2)/2$ being the anomalous $g-$factor),
the spin motion is governed by the Thomas-BMT equation~\cite{Thomas:1927yu,Bargmann:1959gz} (simplified by assuming
$\vec v \cdot \vec B = \vec v \cdot \vec E=0$):
\begin{eqnarray}
\frac{{\rm d} \vec S}{{\rm d}t} &=& \vec S \times \vec \Omega \quad \mbox{with}\nonumber \\
\vec \Omega &=& \frac{e\hbar}{mc} 
\large[ G \vec B +\left(G  - \frac{1}{\gamma^2-1}  \right) \vec E \times \vec v +
\frac{1}{2}{\eta}  \large(  \vec E + \vec v \times \vec B \large) \large] \, .\label{eq:tbmt}
\end{eqnarray}
Here $\vec E$ and $\vec B$ denote the electric and magnetic fields in the
laboratory system.
%$\vec E \cdot \vec v = \vec B \cdot \vec v=0$ has been assumed.
The dimensionless parameter $\eta$ has been introduced via the relation
$\vec d = {\eta} \frac{e \hbar}{2 mc} \vec S$.
The other variables have their usual meaning.

Taking eq. \ref{eq:tbmt} as a starting point, different approaches are possible.
They will be discussed in the following subsections.
In general it is advisable to eliminate the terms proportional to $G$
because spin motions caused by the magnetic moment are in general much larger
than those caused by the tiny EDM effect. 

\subsection{Pure electric field}
Using only an electric field (i.e. $\vec B=0$) with the additional condition that
$\left(G  - \frac{1}{\gamma^2-1}  \right) =0$, 
eq.~\ref{eq:tbmt} reduces to 
\begin{equation}
 \frac{{\rm d} \vec S}{{\rm d}t} = \frac{e \hbar}{2 mc }{\eta}   \vec S \times \vec E \, .
\end{equation}
The condition $\left(G  - \frac{1}{\gamma^2-1}  \right) =0$ 
can only be fulfilled for particles with $G>0$, e.g. for protons with
a  momentum of $p_{\mbox{magic}} =0.7$GeV/$c$.
Using electric fields of the order of about 10 MV/m results in a ring of about 40\,m radius.

Such an all-electric ring is proposed at Brookhaven National Laboratory (BNL)
to measure the EDM of the proton~\cite{bnl}.

\subsection{Combined $E$ and $B$ fields}
With a combined $E$/$B$ ring it is possible to eliminate terms proportional to $G$
if the condition
\[
G \vec B + \left( G - \frac{1}{\gamma^2-1}  \right) \vec E \times \vec v=0
\]
is fulfilled.
This can be achieved for particles with arbitrary $G$.
Such an all-in-one ring is under study by the {\sc JEDI} (J\"ulich Electric Dipole moments Investigations) collaboration
at Forschungszentrum J\"ulich~\cite{jedi}.

\subsection{Pure magnetic field}
In a pure magnetic ring (i.e. $\vec E=0$) eq.~\ref{eq:tbmt} reduces to
\begin{equation}
 \vec \Omega = \frac{e\hbar}{mc} 
\left(G \vec B +
\frac{1}{2}{\eta} \vec v \times \vec B \right) \, .
\end{equation}

\begin{figure}
\includegraphics[width=0.48\textwidth]{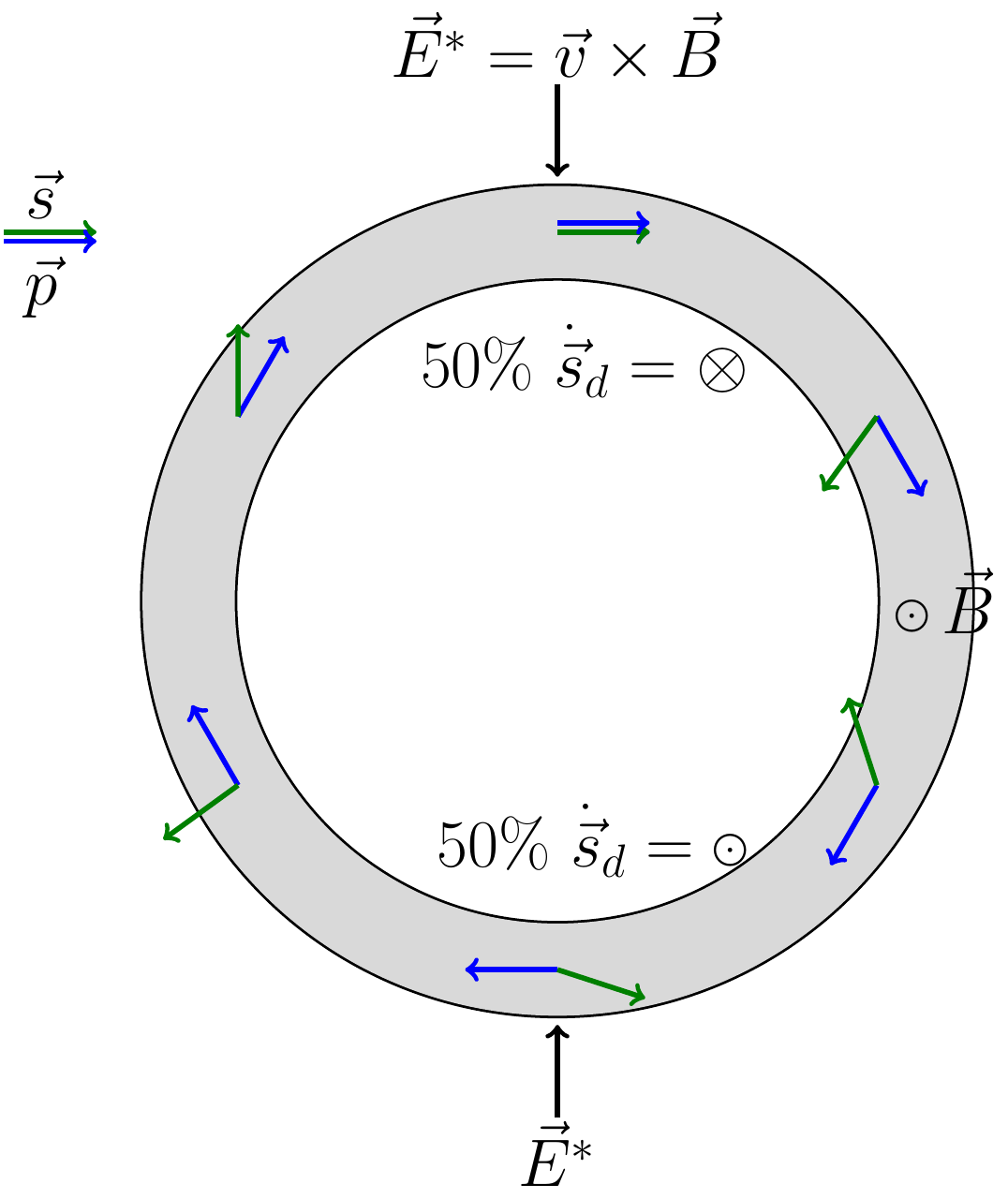}
\includegraphics[width=0.48\textwidth]{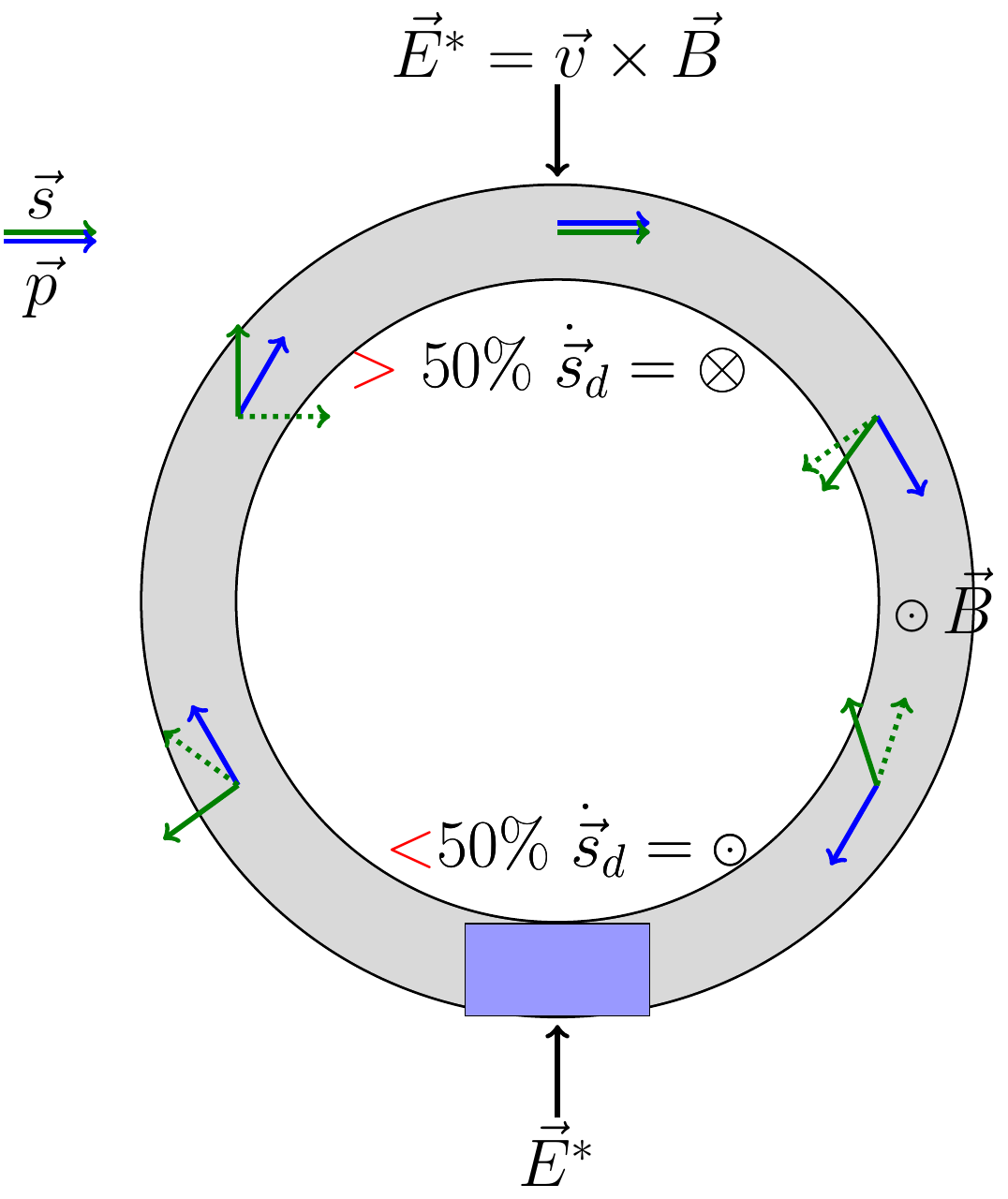}
\caption{Measurement of an EDM with a pure magnetic ring: \newline
Left: In a pure magnetic ring particles feel a radial motional electric field
 $\vec v \times \vec B$. This causes a tilt of the spin vector due to an EDM
out of the plane, e.g. in the upper hemisphere if the spin vector points parallel
to the momentum vector and in the lower hemisphere if it points
anti-parallel. 
This leads to an up-down movement of the spin vector due to an EDM and no
vertical polarization will build up. \newline 
Right: A Wien-filter (blue box) will not affect the particle momentum (for the
reference particle)
but will influence the spin vector (dotted arrows) in such a way that it will point
e.g. more than 50\% parallel to the momentum vector. As a result a vertical
polarization will build up due to an EDM. 
 \label{fig:wien}}

\end{figure}

Here the build-up of an EDM effect is less obvious.
The term proportional to $G$ result in a spin precession in the
horizontal plane of the storage ring.
Due to this precession 50\% of the time the projection of the spin vector
is pointing parallel to the momentum vector and 50\% anti-parallel.
The electric field in the particle's rest frame caused by the
laboratory magnetic field leads thus to an up-down movement of the spin
due to the EDM. No vertical polarization will build up.
Installing a resonant $\vec E/\vec B$ field combination with the condition $\vec E + \vec v \times \vec B =0$ at one or several places in the ring
will have the following effect.
The particle trajectory will not be affected ($\vec E^*=0$) but the spin precession is perturbed ($\vec B^*\ne0$) in such way that the symmetry
between spin parallel and anti-parallel along the momentum vector is broken.
If this so called ``magic Wien filter'' is operated at a correct resonance frequency
given by $f_{\mbox{Wien}} = \left(k + \gamma G \right) f_{\mbox{rev}}$, 
$k$ being an integer and $f_{\mbox{rev}}$ the revolution frequency of the
particles in the storage ring,
a vertical polarization can build up and be measured as an EDM signal (see Fig.~\ref{fig:wien}).
Such an approach is under discussion at Forschungszentrum J\"ulich and could be performed
at the existing (upgraded) storage ring COSY.
A similar method using a radio-frequency radial electric field is 
discussed in~\cite{Lehrach:2012eg}.

%\paragraph{Paragraph headings} Use paragraph headings as needed.

\subsection{Statistical and Systematic accuracy}
The statistical accuracy of the measurement is given by
\begin{equation}\label{stat_err}
\sigma \approx \frac{b \hbar}{\sqrt{N f T \tau_p} P E A } \, .
\end{equation}
$b$ is a factor of the order of 1. 
 Its exact value depends on the details
of the injection and polarization measurement.
The other variables are defined in Tab.~\ref{tab:stat_err}. 
Taking the values given in Tab.~\ref{tab:stat_err} a statistical error of $10^{-29}e\cdot$cm
per year can be reached for a dedicated ring (all-electric, all-in-one ring).
For the pure magnetic ring where the EDM effect is only achieved with help
of the resonant Wien-filter
the sensitivity is of the order of $10^{-24}e\cdot$cm per year.

\begin{center}
\begin{table}[b!]
\begin{tabular}{l|l|l}
variable  & meaning   &  value\\
\hline
 $P$       & beam polarization    & 0.8 \\
$\tau_p$  & Spin coherence time/s & 1000 \\
$E$       & Electric field/MV/m   & 10 \\
$A$       & Analyzing Power    &  0.6 \\
$N$       & nb. of stored particles/cycle  & $4 \times 10^7$ \\
$f$       & detection efficiency   & 0.005\\
$T$       & running time per year/s    & $10^7$\\
\end{tabular}
\caption{Typical values of parameters relevant for the statistical accuracy.\label{tab:stat_err}}
\end{table}
\end{center}

One major source of systematic error is a 
residual radial $B$ field which mimics an EDM effect.
If $\mu B_r \approx d  E_r$,  
a radial magnetic field of  $B_r = \frac{d  E_r}{\mu_N} \approx 3 \, \cdot 10^{-17}$T
causes the same effect as the EDM assuming
$d=10^{-29}e \cdot$cm in a field of $E=10$MV/m. 
To fight such systematic errors the use of two beams
running clock and counter-clockwise is proposed.
A radial field $B_r$ would result  in a vertical separation of the two beams.
For a more detailed discussion on systematic errors see~\cite{bnl,yannis}.

Many more tests and systematic studies are needed and are foreseen 
to reach the target numbers given in Tab.~\ref{tab:stat_err}.
In a first step using correction sextupoles the spin coherence time could already be increased
from a few seconds to about 200~s in the COSY storage ring.

\subsection{Comparison of the different methods}

Tab.~\ref{tab:comp_methods} lists the advantages and disadvantages
of the three complementary approaches discussed.
The first two approaches demand the construction of new dedicated storage rings.
The third approach can be achieved on a much shorter time scale using
the (upgraded) existing storage ring COSY.

\begin{center}
\begin{table}
\begin{tabular}{l|l|l}
                   &  advantage        &  disadvantage \\
\hline
1.) pure electric ring &  no $\vec B$ field needed       &  works only for $p$ \\
$\quad\,$ (BNL)  &  &at fixed momentum \\
\hline
2.) combined ring      &  works for $p,d$, $^3$He, \dots &  both $\vec E$ and $\vec B$ \\
$\quad\,$(J\"ulich)                   &                                 &  required                \\
\hline
3.) pure magnetic ring & existing (upgraded) COSY       &lower sensitivity \\
 $\quad\,$ (J\"ulich)                   & ring can be used,               &      \\
                  & shorter time scale             &                           \\
\hline
\end{tabular}
\caption{Comparison of the various methods discussed.\label{tab:comp_methods}}
\end{table}
\end{center}

%{Main Challenges}
%\begin{itemize}
%\item Spin Coherence Time (SCT)$\approx 1000$s
%\item Beam positioning $\approx$ 10nm (relative between CW-CCW)
%\item Polarimetry on 1 ppm level
%\item Field Gradients $\approx 10$MV/m
%\end{itemize}

\section{Summary \& Outlook}

EDMs of (charged) hadrons are of high interest
to disentangle various sources of $\mathcal{C} \mathcal{P}$ violation
searched for to explain matter - antimatter asymmetry in the Universe. 
A step-wise approach to perform such measurements has been presented.
After investigations of systematic errors at the existing COSY ring
an upgraded COSY storage ring will be used to perform a first direct measurement
of a charged hadron EDM. The next step will be the construction
of dedicated storage rings at  For\-schungs\-zen\-trum J\"ulich in Germany
(all-in-one-ring for proton, deuteron and light nuclei)
and Brookhaven National Laboratory in the USA (all-electric ring for proton) to reach for a higher sensitivity
of $10^{-29}e\cdot $cm.

%\begin{figure}
%  \includegraphics{example.eps}
%\end{figure}
%\begin{figure*}
%  \includegraphics[width=0.75\textwidth]{example.eps}
%\end{figure*}
%\begin{table}
%\begin{tabular}{lll}
%\hline\noalign{\smallskip}
%\end{tabular}
%\end{table}

%\begin{acknowledgements}
%\end{acknowledgements}



%\bibliography{~/mybib}

\end{document}